\documentclass[universe,article,accept,pdftex,moreauthors]{Definitions/mdpi}
\graphicspath{{Figures/}}
\usepackage{multirow}
\firstpage{1} 
\pubvolume{1}
\issuenum{1}
\articlenumber{0}
\pubyear{2024}
\copyrightyear{2024}
\externaleditor{Academic Editors: {Vasiliki Mitsou, Armen Sedrakian and \linebreak David Blaschke}}
\datereceived{10 February 2024}
\daterevised{27 March 2024}
\dateaccepted{8 April 2024}
\datepublished{}
\hreflink{https://doi.org/} 

\usepackage{comment}

\def\apj{Astrophys. J.}
\def\apjl{Astrophys. J. Lett.}

\def\aap{Astron. Astrophys. }

\def\mnras{Mon. Not. R. Astron. Soc. }

\def\prl{Phys. Rev. Lett.}

\newcommand{\be}{\begin{equation}}
\newcommand{\ee}{\end{equation}}
\newcommand{\bea}{\begin{eqnarray}}
\newcommand{\eea}{\end{eqnarray}}

\def\gsim{\mathrel{\raise.5ex\hbox{$>$}\mkern-14mu
\lower0.6ex\hbox{$\sim$}}}
\def\lsim{\mathrel{\raise.3ex\hbox{$<$}\mkern-14mu
\lower0.6ex\hbox{$\sim$}}}

\Title{
A New Solution of the Pulsar Equation}

\TitleCitation{A New Solution of the Pulsar Equation}

\Author{ {Ioannis} 
Contopoulos $^{1,}$ {*}, Ioannis Dimitropoulos $^{1,2}$, Dimitris Ntotsikas $^{2}$ and Konstantinos N. Gourgouliatos $^{2}$}

\AuthorNames{Ioannis Contopoulos, Ioannis Dimitropoulos, Dimitris Ntotsikas,  Konstantinos N. Gourgouliatos}

\AuthorCitation{ {Contopoulos} 
, I.; Dimitropoulos, I.; Ntotsikas, D.; Gourgouliatos, K.N.}

\address{%
$^{1}$ \quad Research Center for Astronomy and Applied Mathematics, Academy of Athens, 11527 Athens, Greece;  {johndim888@gmail.com}\\  
$^{2}$ \quad Department of Physics, University of Patras, 26504 Rio, Greece;  {d.ntotsikas@ac.upatras.gr (D.N.); kngourg@upatras.gr; (K.N.G.)}}

\corres{Correspondence: icontop@academyofathens.gr}

\abstract{We present the first new type of solution of the pulsar equation since 1999. In it, the whole magnetosphere is confined inside the light cylinder and an electrically charged layer wraps around it and holds it together. The reason this new solution has never been obtained before is that all current time-dependent simulations are initialized with a vacuum dipole configuration that extends to infinity; thus, their final steady-state solution also extends to infinity. Under special conditions, such a confined configuration may be attained when the neutron star first forms in the interior of a collapsing star during a supernova explosion, or when it accretes from an external wind or disk from a donor star. It is shown that this new maximally closed non-decelerating solution is the limit of a continuous sequence of standard magnetospheres with open and closed field lines when the amount of open field lines gradually drops to zero. The minimum energy solution in this sequence is a standard magnetosphere in which the closed field line region extends up to about $80\%$ of the light cylinder. We estimate that the released energy when the new solution transitions to the minimum energy one is enough to power a fast radio burst.}


\begin{document}

\setcounter{section}{0} 


\section{The Pulsar Equation}

{
Right after the discovery of pulsars in 1968~\cite{hewish1969observation}, Goldreich and Julian were the first to sketch the general structure of the pulsar magnetosphere~\cite{goldreich1969pulsar}. The detailed structure of the axisymmetric pulsar magnetosphere was obtained much later by Contopoulos, Kazanas and Fendt, who clearly identified the presence of a global magnetospheric poloidal electric current sheet~\cite{CKF99}. Since then, several authors have addressed the pulsar magnetosphere problem with force-free electrodynamic (FFE) simulations (\cite{S06,kalapotharakos2009three}, etc.) magnetohydrodynamic (MHD) simulations (\cite{tchekhovskoy2013time}, etc.) ‘ab initio’ particle-in-cell (PIC) simulations
(\cite{philippov2014ab,philippov2015ab,kalapotharakos2018three}, etc.) and most recently with Machine Learning~\cite{stefanou2023modelling,Dimitropoulos:2024}. Several inconsistencies in recent state-of-the-art numerical simulations (e.g., the extent of the co-rotating closed-line region, the microscopic treatment and the thickness of the current sheet, magnetospheric dissipation, etc.) led us to believe that one can only trust them qualitatively (not quantitatively) to make meaningful comparisons with observations. This is why, in~\cite{Dimitropoulos:2024}, we proposed to return to the basics and obtain the reference ideal steady-state force-free magnetosphere in a novel independent way, namely with Machine Learning.

The magnetospheres of neutron stars are dominated by the electric and magnetic fields. The physical conditions allow us to neglect gravity, thermal pressure and particle inertia because they are several orders of magnitude smaller than the electromagnetic forces. Therefore, force balance in the bulk of the pulsar magnetosphere is reduced to
\begin{equation}
\rho_e {\bf E} + {\bf J}\times {\bf B}/c\approx {\bf 0}\ ,
\label{forcebalance}
\end{equation} 
where  {${\bf E}$} and  {$\bf{B}$} are the electric and magnetic fields, respectively, $\rho_e=\nabla\cdot {\bf E}/(4\pi)$ is the electric charge density, and ${\bf J}$ is the electric current density. Under axisymmetric conditions, one may define the magnetic flux function $\Psi$ such that
 {${\rm B}_r$} $=\partial\Psi/\partial\theta/(r^2\sin\theta)$ and ${\rm B}_\theta=-\partial\Psi/\partial r/(r\sin\theta)$, which automatically satisfies the condition $\nabla\cdot{\bf B}=0$. 
}
Under steady-state axisymmetric force-free ideal conditions, Equation~(\ref{forcebalance}) becomes the well-known pulsar equation, namely
\begin{adjustwidth}{-\extralength}{0cm}
\begin{eqnarray}
\left(1-\frac{r^2 \sin^2\theta}{R_{\rm LC}^2}\right) \left[ \frac{\partial^2 \Psi}{\partial r^2}-\frac{\partial \Psi}{\partial\theta}\frac{\cos\theta}{r^2 \sin\theta}+\frac{1}{r^2}\frac{\partial^2 \Psi}{\partial\theta^2}\right]-\frac{2r\sin\theta}{R_{\rm LC}^2}\left[\frac{\partial\Psi}{\partial\theta}\frac{\cos\theta}{r}+\frac{\partial \Psi}{\partial r} \sin\theta\right] +II'(\Psi)=0  \label{pulsareq1}
\end{eqnarray}
\end{adjustwidth}
in spherical coordinates $r,\theta$~\cite{scharlemann1973aligned}. Here, $I=I(\Psi)$ is the distribution of poloidal electric current from which one obtains the azimuthal component of the magnetic field ${\rm B}_\phi=I/(r\sin\theta)$, and $I'\equiv {\rm d}I/{\rm d}\Psi$. Equation~(\ref{pulsareq1}) is an elliptical partial differential equation with a singularity along the light cylinder, the nominal distance where the co-rotational velocity becomes equal to the speed of light ($R_{\rm LC}\equiv c/\Omega$, where $\Omega$ is the angular velocity of stellar rotation). This equation was first solved with a special numerical technique developed by~\cite{CKF99} that was used again since then by several others (e.g.,~\cite{Gruzinov:2005, Timokhin:2006, Ntotsikas:2024}).

In  {Section}~\ref{sec:2},  
we present a new type of solution of Equation~(\ref{pulsareq1}) that is very different from the canonical solutions found in the literature. In  {Section}~\ref{sec:3}, we obtain new solutions of the ideal force-free pulsar magnetosphere for various amounts of open magnetic flux using the novel Machine Learning methodology of~\cite{Dimitropoulos:2024} (hereafter Paper~I; see description in Appendix~\ref{app:level1}). We show that the new type of solution is the limit of a continuous sequence of standard magnetospheres with open and closed field lines in which the closed line region approaches closer and closer to the light cylinder and the amount of open field lines gradually drops to zero. In  {Section}~\ref{sec:4} we discuss the possibility that an abrupt transition from our new magnetospheric solution to a standard magnetosphere with open and closed field lines may generate a fast radio burst. We conclude in  {Section}~\ref{sec:5} with a discussion of our results. We consider only axisymmetric magnetospheres, and in future work, we will investigate whether our results apply also to an oblique rotator. 

\section{A Confined Magnetosphere}\label{sec:2}

\begin{figure}[H]
\includegraphics[width=10cm,angle=0]{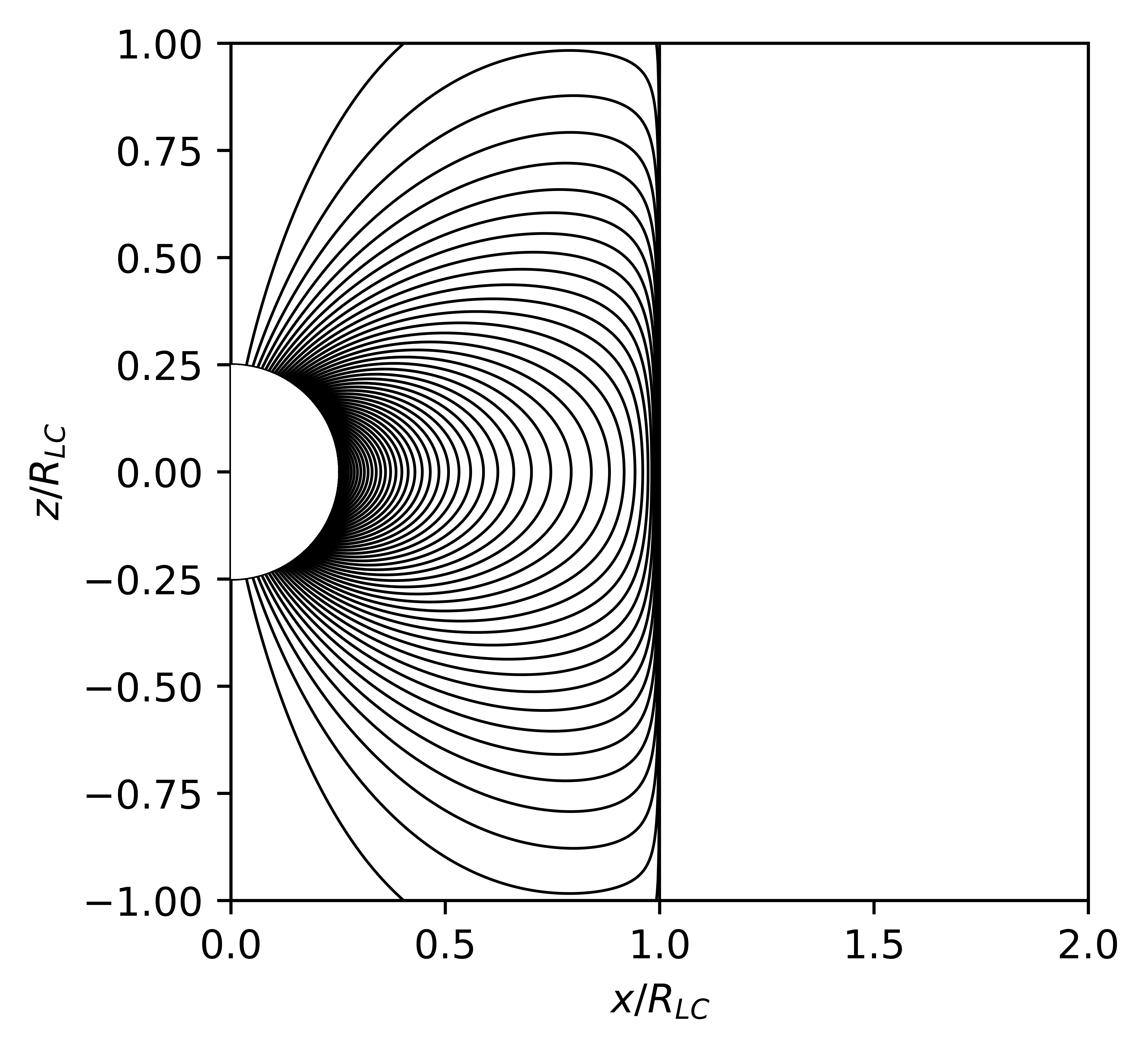}
\caption{New co-rotating confined magnetospheric solution. $\Psi=0$ along the axis $\theta=0,\pi$, $z=\pm\infty$, and the light cylinder $R=R_{\rm LC}$. $\Psi$ contour levels: integer multiples of $0.1\Psi_{\rm dipole\ LC}$ with $\Psi=0$ along the $z$-axis. Thick vertical line: charged current sheet along the light cylinder.}
\label{crazymag}
\end{figure}

Let us consider here a theoretical limit that has never been considered before, namely one with no open field lines and no poloidal electric current ($I=0$). In that limit, the whole magnetosphere of the central dipole is contained within the light cylinder,
with boundary conditions $\Psi(r,\theta=0)=\Psi(r,\theta=\pi)=0$, a dipole field $\Psi(r_*,\theta)=\Psi_{\rm max}\sin^2\theta$ along the surface of the central star ($\Psi_{\rm max}$ is the total dipolar magnetic flux that emanates from each stellar hemisphere), and $\Psi(r=R_{\rm LC}/\sin\theta,\theta)=0$ along the light cylinder. The solution of Equation~(\ref{pulsareq1}) with these boundary conditions is obtained with the method described in~\cite{Gourgouliatos:2008} and is shown in Figure~\ref{crazymag}.
This configuration co-rotates with the central star and is wrapped around by an electrically charged layer at the light cylinder. It is a unique feature of special relativity that allows such a spatially confined electromagnetic field solution, namely the presence of electric charges and electric fields that are generated by the rigid co-rotation of the magnetosphere. In the region next to the light cylinder in particular, the electric charge density is positive/negative and generates (via its co-rotation) an azimuthal electric current along/opposite to the direction of rotation for an aligned/counter-aligned rotator, respectively. The magnetic field points downwards/upwards and the electric field outwards/inwards, respectively, so it is obvious that the inward Lorentz force ${\bf J}\times {\bf B}/c$ is balanced by the outward electrostatic force $\rho_e {\bf E}$.

It is, however, interesting that such a confined solution with a vacuum outside is not possible without a light cylinder, as is the case in non-relativistic MHD. In particular, pressure balance across the electrically charged current sheet along the light cylinder may be written as
\begin{equation}
\left. ({\rm B}^2-{\rm E}^2)\right|_{\rm IN}=\left. ({\rm B}^2-{\rm E}^2)\right|_{\rm OUT}=0\ ,
\label{pressurebalance1}
\end{equation}
or equivalently
\begin{equation}
\left. {\rm B}^2_z(1-x^2)\right|_{\rm IN}=0\ .
\label{pressurebalanceLC}
\end{equation}

The requirement for the continuity of ${\rm B}^2-{\rm E}^2$ across an infinitely thin relativistic current sheet, Equation~(\ref{pressurebalance1}), stems from the integration of the force-free equation $\rho_e{\bf E}+{\bf J}\times{\bf B}/c={\bf 0}$ across it~\cite{lyubarsky2005relativistic,uzdensky2003axisymmetric}. Here, $x\equiv R/R_{\rm LC}$ is the cylindrical radius in units of the light cylinder radius $R_{\rm LC}$, and $\left. x\right|_{\rm IN}\rightarrow 1^-$. It is assumed that the whole magnetosphere is confined inside the light cylinder, it remains untwisted (i.e., ${\rm B}_\phi=0$), it co-rotates with the central star, and there is nothing outside. Obviously, in the limit $\left. x\right|_{\rm IN}=1^-$, pressure balance (i.e.,  Equation~(\ref{pressurebalanceLC})) is satisfied; thus, solutions spatially confined inside the light cylinder are possible.  In this new solution, the value of $\left. {\rm B}_z\right|_{\rm IN}$ is non-zero and can be determined numerically. We emphasize once again that such spatially self-confined solutions with a vacuum outside are possible only along the light cylinder, not inside or outside.

\begin{figure}[H]
\includegraphics[width=4.5cm,angle=0]{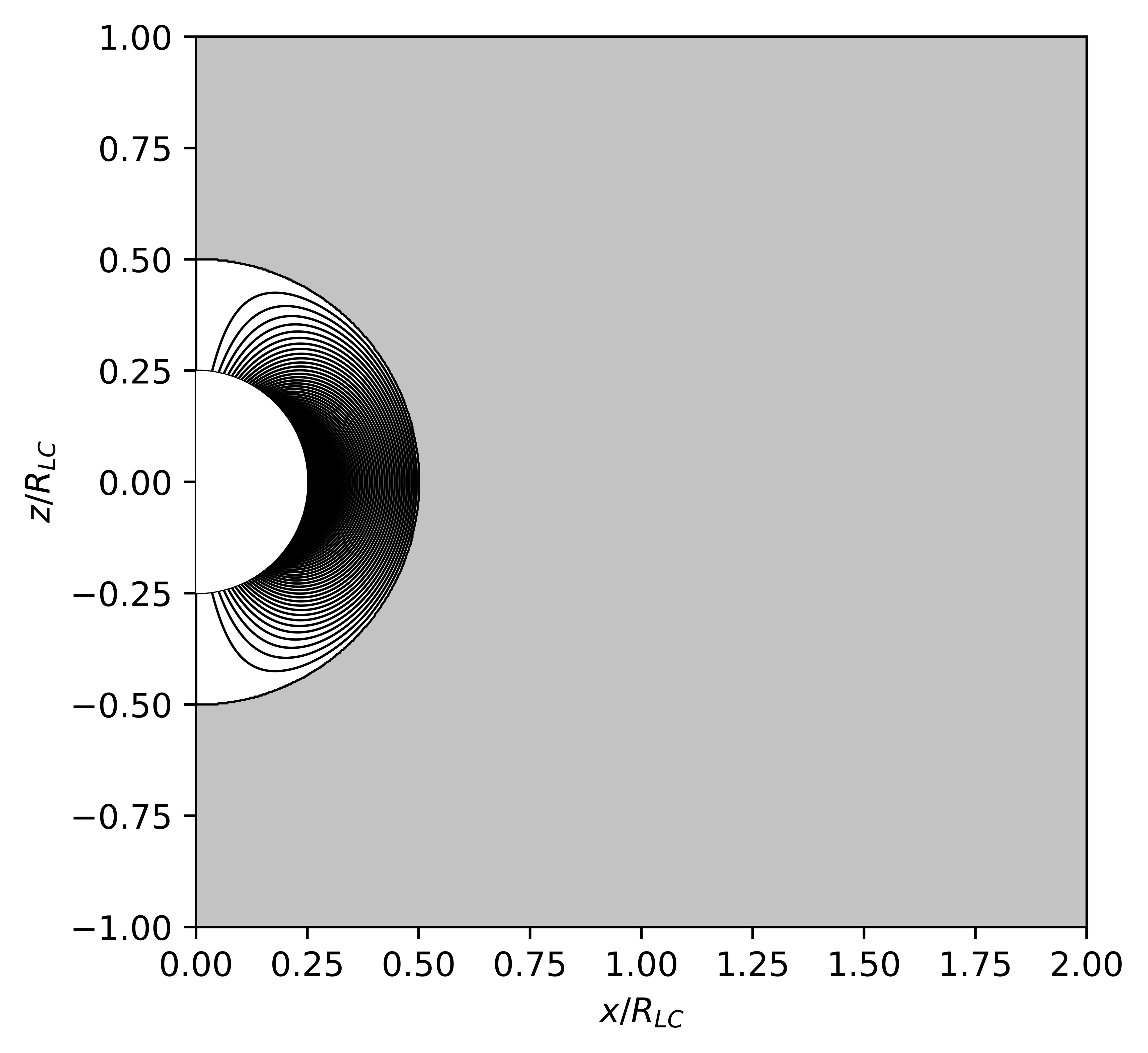}
\hspace{-0.3cm}
\includegraphics[width=4.5cm,angle=0]{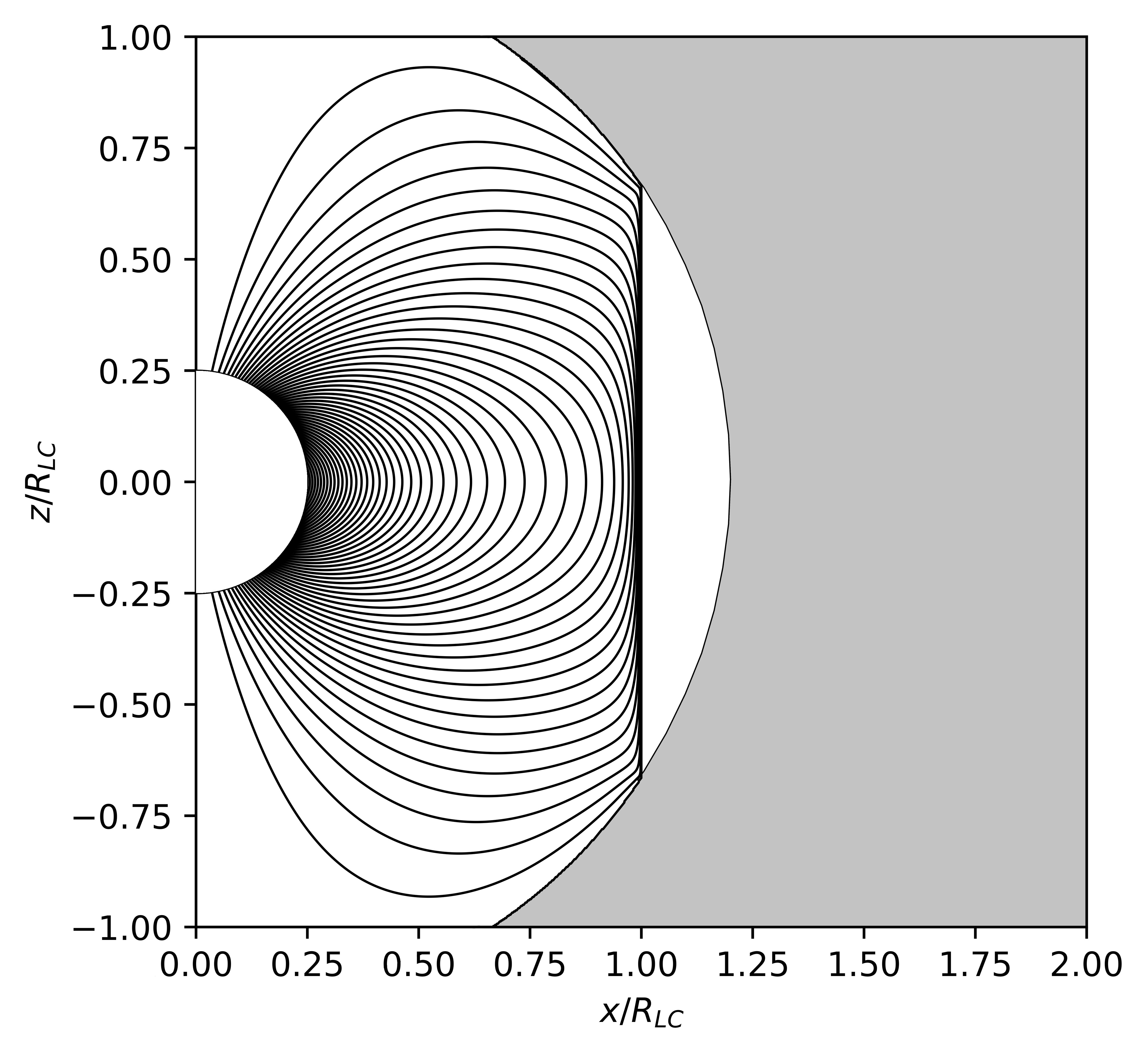}
\hspace{-0.3cm}
\includegraphics[width=4.5cm,angle=0]{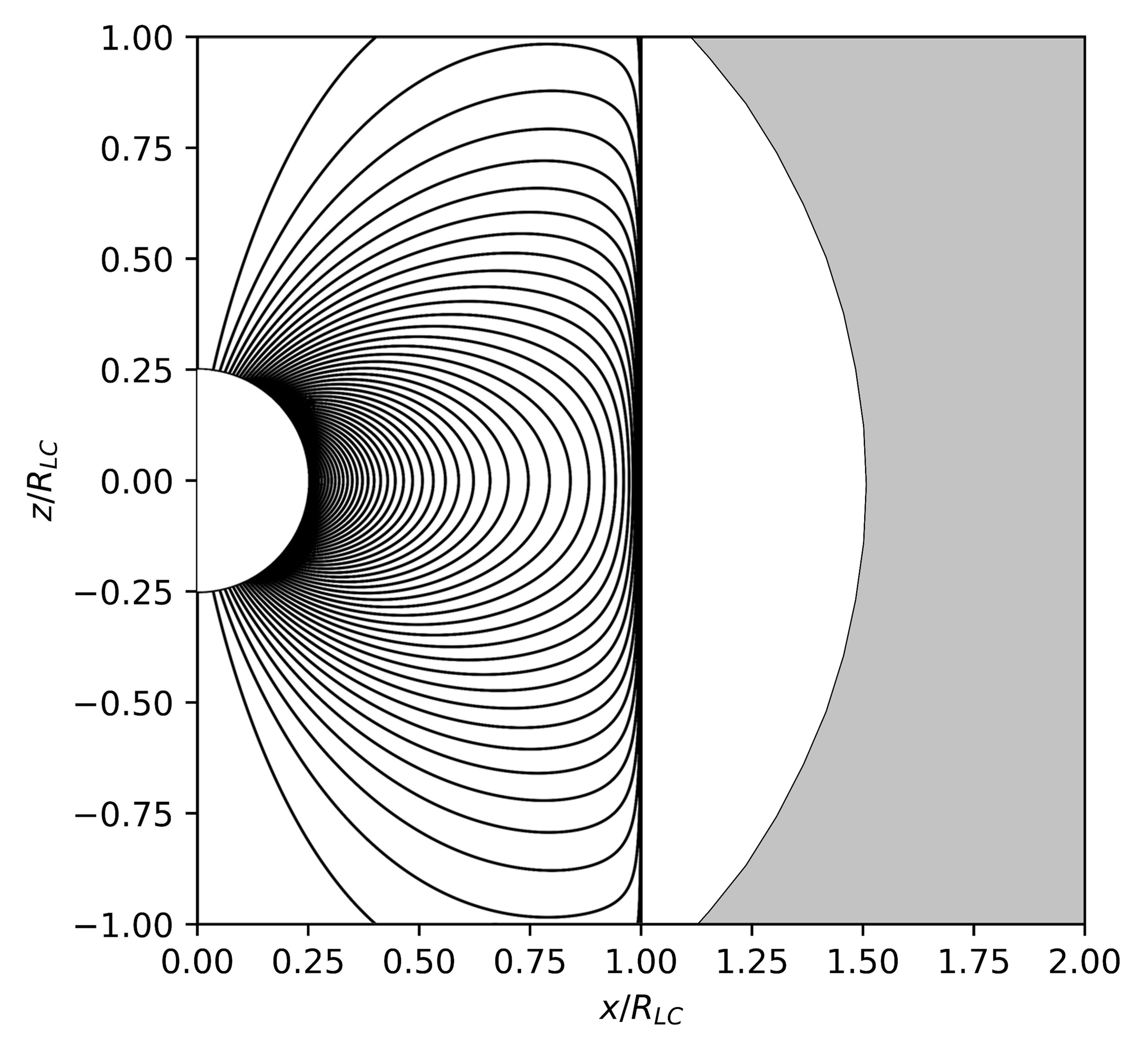}
\caption{ {Sequence} of co-rotating magnetospheric solutions confined by an expanding shell of plasma material (grey region). As the ejecta expand to infinity, the solution approaches the solution shown in Figure~\ref{crazymag}.}
\label{crazymag2} 
\end{figure}

We propose that this singular configuration may be realized in nature under special conditions during the stellar collapse that led to the formation of a spinning neutron star in a supernova explosion. The initial neutron star magnetosphere will be squeezed to a thin layer above the surface of the neutron star and will rotate with the neutron star.
The solutions of the pulsar equation shown in Figure~\ref{crazymag2} were obtained with the same method by setting $\Psi=0$ along a sphere of radius $r_{\rm ejecta}$, and along the light cylinder wherever $r_{\rm ejecta}\sin\theta>R_{\rm LC}$. Thus, the whole magnetosphere is confined between the neutron star surface and a spherical shell of supernova ejecta. As the supernova explosion proceeds and the ejecta expand, the size $r_{\rm ejecta}$ of the magnetosphere will increase. During that stage, the whole magnetosphere co-rotates with the central neutron star; thus, it will be confined inside the light cylinder wherever $r_{\rm ejecta}\sin\theta>R_{\rm LC}$. It is implied here that, during the supernova explosion, as $r_{\rm ejecta}$ grows slowly at subluminal velocities, the magnetosphere evolves fast into a sequence of steady-state solutions. When $r_{\rm ejecta}$ grows beyond $r_{\rm ejecta}=R_{\rm LC}$, any part of the magnetosphere that crosses the light cylinder cannot open up to infinity as in the magnetosphere of an isolated pulsar. Therefore, it will be wound very fast by the stellar rotation, the Lorentz force will squeeze it vertically, and it will eventually detach from the rest of the magnetosphere that remains confined inside the light cylinder as in the solutions shown in Figure~\ref{crazymag2}. 
{
It is obvious that this is an idealized situation in which a neutron star and its magnetosphere formed and remained at the center of the collapse (in general, the crust of a young neutron star may  not have formed/solidified yet, and the neutron star would also potentially experience a kick that will move it away from the center of the collapse). It is also conceivable that such a confined configuration is attained when a neutron star accretes from a surrounding disk or from external winds from a donor star (e.g.,~\cite{Shakura:1973, Doroshenko:2011}). It is also obvious that the solutions shown in Figures~\ref{crazymag} and \ref{crazymag2} are force-free approximations since they were obtained without taking into consideration the baryon-rich and dense environment of a newborn neutron star.
}

{
The idealized solution shown in Figure~\ref{crazymag}
is most probably unstable. We have not performed a formal stability analysis of the light cylinder boundary, but we understand this instability as follows:
Whenever a part of the confined magnetosphere extends beyond the light cylinder, it cannot keep co-rotating with the central star without an azimuthal component of the magnetic field. In other words, when an initially poloidal magnetic field line crosses the light cylinder, it is swept back so that the particle drift velocity 
\begin{eqnarray}
v_{\rm D}&\equiv& \frac{|{\bf E}\times {\bf B}|}{{\rm B}^2}c=
\frac{|{\bf E}_p\times {\bf B}_p+{\bf E}_p\times {\rm B}_\phi \hat{\phi}|}{{\rm B}_p^2+{\rm B}_\phi^2}c\nonumber\\
&=&
\frac{\sqrt{{\rm B}_p^4 x^2 + {\rm B}_p^2 {\rm B}_\phi^2 x^2}}{{\rm B}_p^2+{\rm B}_\phi^2}c=
\frac{{\rm B}_p x}{\sqrt{{\rm B}_p^2+{\rm B}_\phi^2}}c<c
\label{vD}
\end{eqnarray}
always remains subluminal. Without the azimuthal component ${\rm B}_\phi$, $v_D>c$ outside the light cylinder, where $x>1$.
Therefore, any part of the confined magnetosphere that crosses the light cylinder will develop an azimuthal magnetic field component that will push it outwards toward infinity. Thus, the magnetosphere will transition to a standard solution with open and closed field lines, emitting an electromagnetic pulse of azimuthal magnetic field along the way. A similar effect is seen in numerical simulations during the evolution of initially poloidal magnetic field configurations when the stellar rotation is initiated (e.g.,~\cite{S06}). We will return to a further discussion of this burst in Section~\ref{sec:3} below.
}
{
 {Notice that using} Equation~(\ref{vD}) with ${\rm B}_\phi=0$, the drift velocity is exactly equal to $c$ at the light cylinder, and the neglect of inertial terms is more questionable for the confined solution than for the standard one near the light cylinder. Inertial effects are even more pronounced near the light cylinder because, due to the $(1-x^2)$ term in the denominator of the Goldreich--Julian charge density $\rho_e=-{\rm B}_z/[2\pi R_{\rm LC}(1-x^2)]$~\cite{Timokhin:2006}, the volume integral of $\rho_e$ and the corresponding mass content of the idealized confined magnetosphere diverge.  {This suggests that, if realized in nature, the confined solution will be limited to within some short distance from the light cylinder.}
} 

\section{A Spectrum of Solutions}\label{sec:3}

We will now show that the new solution obtained in the previous section is the limit of a continuous spectrum of solutions of the standard type, namely solutions that consist of a co-rotating closed-line region and an open-line region that extends to infinity.
{
Figure~\ref{crazymag} is the mathematical limit of a sequence of independent standard solutions, and does not imply a physical evolution of standard solutions. Nevertheless, each solution contains a different amount of electromagnetic energy, so magnetospheric transitions from a higher to a lower energy solution are possible (see below). 
}

\begin{figure}[H]
\includegraphics[width=6.4cm,angle=0.0]{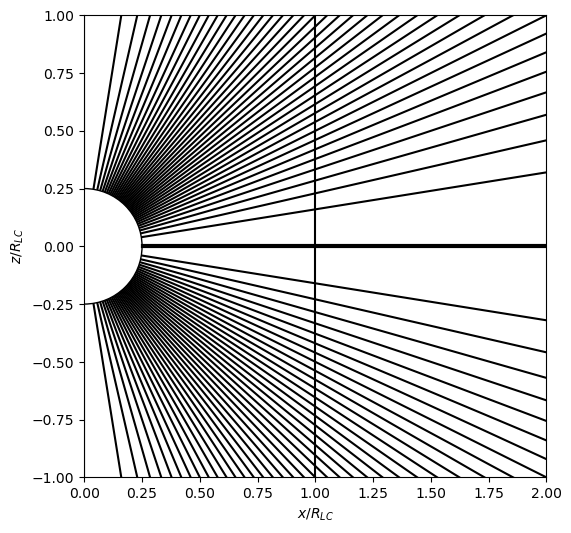}
\hspace{-0.3cm}
\includegraphics[width=6.4cm,angle=0.0]{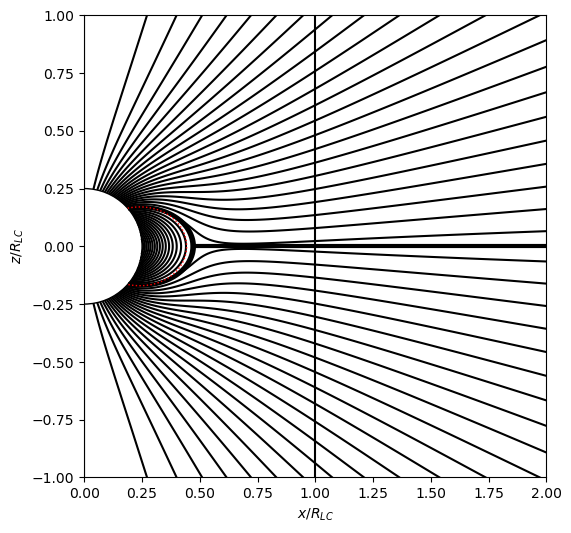}
\hspace{-0.3cm}
\includegraphics[width=6.4cm,angle=0.0]{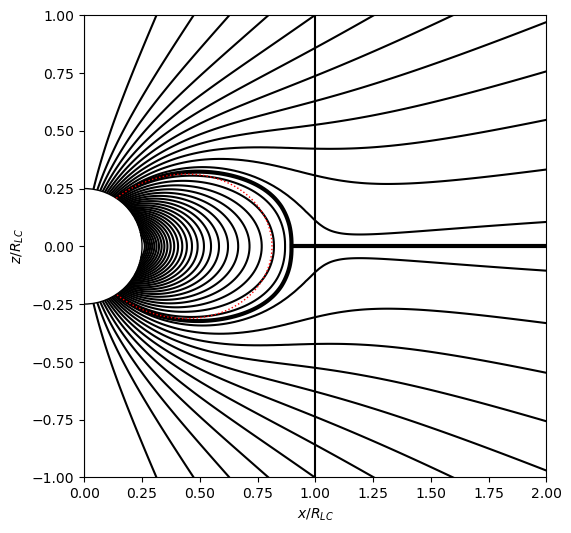}
\hspace{-0.3cm}
\includegraphics[width=6.4cm,angle=0.0]{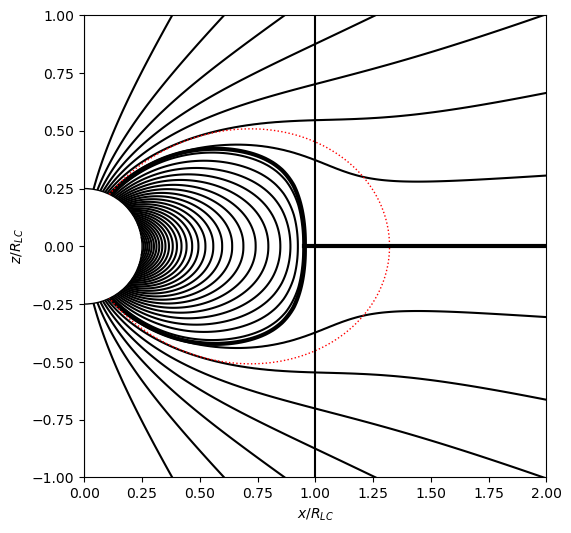}
\includegraphics[width=6.4cm,angle=0.0]{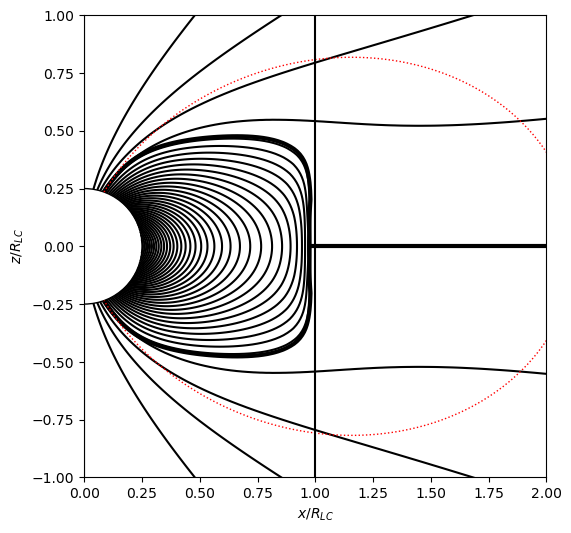}
\hspace{-0.3cm}
\includegraphics[width=6.4cm,angle=0.0]{figure1.png}
\caption{ {Intermediate} solutions for $\theta_{\rm pc}=\pi/2$ and $\theta_{\rm pc}=2/1.176/0.9/0.7/0\times (r_*/R_{\rm LC})^{1/2}$ (top left/top right/middle left/middle right/bottom left/bottom right, respectively), or equivalently $\Psi_{\rm open}=\Psi_{\rm max}$ and $\Psi_{\rm open}\equiv \Psi_{\rm S}=\Psi_{\rm max}\sin^2\theta_{\rm pc}=2.256/1.228/0.756/0.468/0\times \Psi_{\rm dipole\ LC}$, respectively. $\Psi$ contour levels: integer multiples of $0.1\Psi_{\rm dipole\ LC}$ with $\Psi=0$ along the $z$-axis. Red dotted lines: initial dipolar shape of the separatrix between open and closed field lines. Thin vertical line: line cylinder. Thick lines: current sheets along the separatrix and equator (and along the light cylinder in the bottom right solution).}
\label{solutions} 
\end{figure}

We obtain here magnetospheric solutions with the methodology of Paper~I (see Appendix~\ref{app:level1}) in which the separatrix between open and closed field lines is considered to be a mathematical contact discontinuity, namely a surface of zero thickness. The tip of the closed-line region is called the Y-point. We consider in particular Y-point positions from $x_{\rm Y}\approx r_*/R_{\rm LC}\ll1$ to $x_{\rm Y}\approx 1$ very close to the light cylinder. For reasons of computational convenience,
we chose a rather large star  with radius $r_*=0.25 R_{\rm LC}$
{
(PINN methods cannot handle well large scale differences in their domain of application).
}
{
We have also tried different (smaller) values of $r_*$ and obtained similar results.
}
As is described in Appendix~\ref{app:level1}, we first choose the angular opening of the polar cap $\theta_{\rm pc}$ from which a certain amount of magnetic flux $\Psi_{\rm open}\equiv \Psi_{\rm S}=\Psi_{\rm max}\sin^2\theta_{\rm pc}$ emanates and extends to infinity. This forms the so-called open-line region. In the solutions shown in \mbox{Figure~\ref{solutions}}, one particular value of $\theta_{\rm pc}$, namely $\theta_{\rm pc}=1.176(r_*/R_{\rm LC})^{1/2}$, was specially chosen with the following in mind: it is straightforward to calculate that, in a dipolar magnetic field configuration, the magnetic field line that crosses the light cylinder corresponds to $\Psi_{\rm dipole\ LC}=\Psi_{\rm max}r_*/R_{\rm LC}=B_* r_*^3 \Omega/(2c)$ (here, $B_*$ is the polar value of the surface magnetic field), and $\theta_{\rm dipole\ pc}=\sin^{-1}[(r_*/R_{\rm LC})^{1/2}]\approx (r_*/R_{\rm LC})^{1/2}$. In previous high-resolution solutions of the pulsar equation (e.g.,~\cite{Timokhin:2006}), the magnetic field line that crosses the light cylinder corresponds to $\Psi_{\rm LC\ Timokhin}=1.23 \Psi_{\rm dipole\ LC}$ and $\theta_{\rm pc\ Timokhin}\approx 1.176(r_*/R_{\rm LC})^{1/2}$. Therefore, this particular value of $\theta_{\rm pc}$ was chosen so that the Y-point lies very close to the light cylinder.  Instead, in Paper~I, it was found that $R_{\rm Y}=0.88 R_{\rm LC}$, which is closer to  $R_{\rm dipole}(\Psi_{\rm S})=r_*/\sin^2\theta_{\rm pc}= 0.81R_{\rm LC}$. This unexpected result is due to our clear treatment of the separatrix surface as a surface of zero thickness which allows the manifestation of the T-point at its tip (see Appendix~\ref{app:level2}). This is not the case in all previous MHD, FFE, and PIC simulations where the separatrix surface has an unphysically large finite thickness which at its tip leads to an exterior Y-point, and a T-point some distance inside it (see Figure~3 of~\cite{Contopoulos:2024} for details).

We also obtained the solution for $\Psi_{\rm open}=0.866\Psi_{\rm dipole\ LC}$ that corresponds to $\theta_{\rm pc}=0.97(r_*/R_{\rm LC})^{1/2}$ and $R_{\rm Y}= 0.94R_{\rm LC}$. The reader can check that this value of  $\Psi_{\rm open}$ yields a pulsar spindown rate equal to $\dot{E}= 0.75\dot{E}_{\rm vacuum}(90^\circ)$. Here, $\dot{E}_{\rm vacuum}(\lambda)$ is the spindown rate of a vacuum dipole rotator with inclination angle $\lambda$, and
$\dot{E}_{\rm vacuum}(90^\circ)=B_*^2 r_*^6 \Omega^4/(6 c^3)\equiv 2\Psi_{\rm dipole\ LC}^2\Omega^2 /(3c)$. One may tentatively generalize our result for non-zero pulsar inclination angles according to~\cite{S06} as $\dot{E}(\lambda)\approx 0.75\dot{E}_{\rm vacuum}(90^\circ)(1+\sin^2\lambda)$, and since $\dot{E}_{\rm vacuum}(\lambda)=\dot{E}_{\rm vacuum}(90^\circ)\sin^2\lambda$, we obtain that
\begin{equation}
\frac{\dot{E}(\lambda)}{\dot{E}_{\rm vacuum}(\lambda)}\approx 0.75\ \frac{1+\sin^2\lambda}{\sin^2\lambda}\geq 1.5\ .
\label{intermittent}
\end{equation}

It is interesting that in all previous solutions of the FFE pulsar magnetosphere, the above ratio was found to be greater than 3 (e.g.,~\cite{li2012resistive}). This value is significantly larger than the ratio of spindown rates $\dot{E}_{\rm ON}/\dot{E}_{\rm OFF}$ observed in the intermittent pulsars PSR B1931+24, PSR J1832+0029 and  PSR J1841-0500 for their corresponding `ON' and `OFF' states (1.5, 1.7 and 2.5, respectively, e.g.,~\cite{rea2008nature,wang2020radio}). The inability to account for observed values lower than 3 was the reason that led to the development of resistive magnetospheric solutions (e.g.,~\cite{kalapotharakos2012toward,li2012resistive}). With our new solutions, it seems that there is no need for magnetospheric resistivity anymore. This result certainly merits further investigation.

We were able to apply our method to even smaller values of $\Psi_{\rm open}$ and $\theta_{\rm pc}$. These solutions are very hard to obtain with the methodology of Paper~I because in these, the Y-point approaches very close to the light cylinder and the convergence of the solution in the open-line region between the Y-point and the light cylinder is very difficult. In the limit $\Psi_{\rm open}=\theta_{\rm pc}=0$, one reaches the solution shown in Figure~\ref{crazymag}, which was obtained independently as a solution of the pulsar equation (eq.~\ref{pulsareq1}) with standard methods (see~\cite{Gourgouliatos:2008}). This spectrum of solutions with smaller and smaller spindown rates is new and has never been obtained before in previous MHD, FFE, nor PIC simulations. Our results are summarized in Table~\ref{Table}.

\begin{table}[H]
\caption{Summary of solutions. $\Psi_{\rm open}$ in units of $\Psi_{\rm dipole\ LC}$. $\dot{E}$ in units of $\dot{E}_{\rm vacuum}(90^\circ)$.}
\label{Table}
\centering
\setlength{\tabcolsep}{1.8mm}
\begin{tabular}{ccccccl}
\toprule
{Solution} & \boldmath{$\theta_{\rm pc}$} &
\boldmath{$\Psi_{\rm open}$} &
\boldmath{ $x_{\rm Y}\Psi_{\rm open}$} &
\boldmath{ $x_{\rm Y}$} &\boldmath{$\dot{E}$} & 
\\  \midrule
1 & $\pi/2$ & $\frac{\Psi_{\rm max}}{\Psi_{\rm dipole\ LC}}$ &  $\frac{r_* \Psi_{\rm max}}{R_{\rm LC}\Psi_{\rm dipole\ LC}}$ & $\frac{r_*}{R_{\rm LC}}$ & $\left(\frac{\Psi_{\rm max}}{\Psi_{\rm dipole\ LC}}\right)^2$ & Split monopole \\ \\
2 & $1.7(\frac{r_*}{R_{\rm LC}})^{\frac{1}{2}}$ & 2.26 & 1.10 & 0.47 & 5.1 & \\
3 & $1.318(\frac{r_*}{R_{\rm LC}})^{\frac{1}{2}}$ & 1.5 & 1.10 & 0.71 & 2.2 & H\&B\ 2022\\
4 & $1.176(\frac{r_*}{R_{\rm LC}})^{\frac{1}{2}}$ & 1.23 & 1.10 & 0.88 & 1.5 & Paper~I \\
5 & $0.9(\frac{r_*}{R_{\rm LC}})^{\frac{1}{2}}$ & 0.76 & 0.72 & 0.96 & 0.6 & \\
6 & $0.7(\frac{r_*}{R_{\rm LC}})^{\frac{1}{2}}$ & 0.47 & 0.46 & 0.98 & 0.2 & \\
7 & 0 & 0 & 0 & 1 & 0 & Figure~\ref{crazymag} \\
\bottomrule
\end{tabular}
\end{table}

\begin{figure}[H]
\includegraphics[width=10cm,angle=0.0]{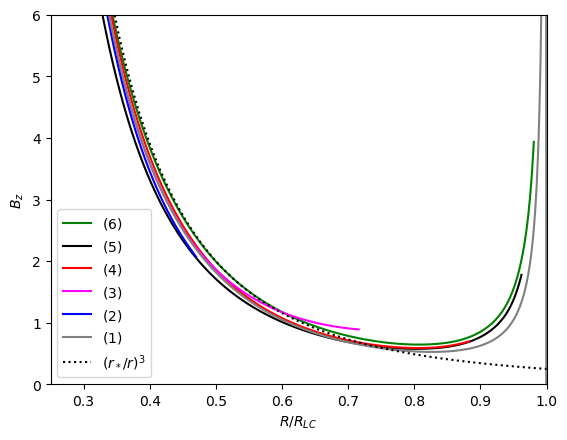}
\caption{The distribution of ${\rm B}_z(r,\theta=\pi/2)$ inside the Y-point for the various solutions listed in Table~\ref{Table}. ${\rm B}_z$ is in arbitrary units.
The dotted line represents the standard $1/r^3$ dipolar dependence. One clearly infers both the outward stretching of the initially dipolar field (at small $r$ distances the solid lines lie below the dotted one), and the inwards squeezing of the field right behind the Y-point that is required to satisfy pressure balance. Never before has anyone seen such a large increase in ${\rm B}_z$ inside the Y-point, even in the solutions in the literature where $x_{\rm Y}\rightarrow 1$. Our results improve Figure~11 of~\cite{Timokhin:2006}.}
\label{Bz}
\end{figure}

In Figure~\ref{Bz}, the distribution of ${\rm B}_z$ is plotted along the equator inside the light cylinder and shows its clear increase right inside the Y-point at its tip, as is needed in order to satisfy Equation~(\ref{div}). Although this effect was known, it has never been observed so clearly before, and Figure~\ref{Bz} improves Figure~11 of~\cite{Timokhin:2006}. 
It is now seen clearly that, with the improved methodology of~\cite{Dimitropoulos:2024}, the minimum ${\rm B}_z$ value of this sequence of solutions lies at around $x_{\rm Y}\sim 0.8$, which is near the position of the Y-point in the high-resolution PIC simulations of~\cite{cerutti2015particle,hu2022axisymmetric,hakobyan2023magnetic,bransgrove2023radio}. 
It may be that the Y-point naturally relaxes to that position (not at the light cylinder), as this corresponds to a minimum energy solution. This result may answer an important question in the recent literature on the pulsar magnetosphere.
{
Note, however, that the numerical experiments in~\cite{kalapotharakos2023gamma} lead to a different interpretation, namely that the Y-point can be pushed arbitrarily close to the stellar surface depending on the number of PIC particles that are either injected or reach the dissipative region beyond the Y-point. This is an interesting interpretation since it suggests that $x_{\rm Y}$ may decrease as the magnetosphere evolves with pulsar spindown. This will lead to a value of the braking index smaller than its canonical value of 3. Otherwise, if $x_{\rm Y}$ remains unchanged, all solutions obtained in Table~\ref{Table} will have braking indices equal to 3.
}

\begin{figure}[H]
\includegraphics[width=10cm,angle=0.0]{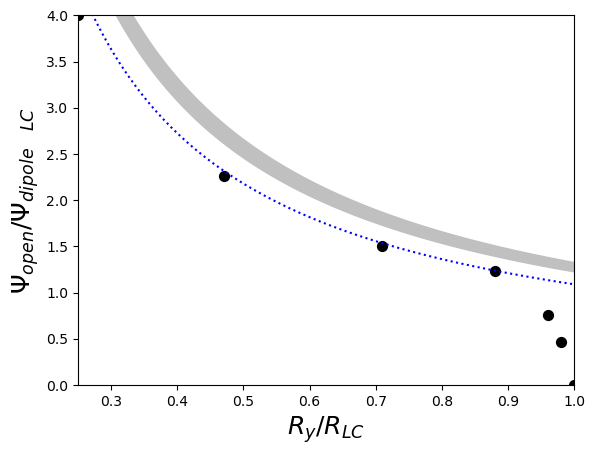}
\caption{ {Open} magnetic flux $\Psi_{\rm open}$ as function of the position of the Y-point $x_{\rm Y}$. Bullet points: solutions listed in Table~\ref{Table}. Dotted blue line: $\Psi_{\rm open}= 1.10 \Psi_{{\rm dipole}\ x_{\rm Y}}=1.10\ \Psi_{\rm dipole\ LC}/x_{\rm Y}$. Grey band: canonical values $\Psi_{\rm open}= (1.23-1.33)\ \Psi_{\rm dipole\ LC}/x_{\rm Y}$ (\cite{CKF99,contopoulos2005coughing,Timokhin:2006,hu2022axisymmetric}, etc.). As $x_{\rm Y}\rightarrow 1$, $\Psi_{\rm open}\rightarrow 0$.}
\label{Psiopen} 
\end{figure}

\begin{figure}[H]
\includegraphics[width=10cm,angle=0.0]{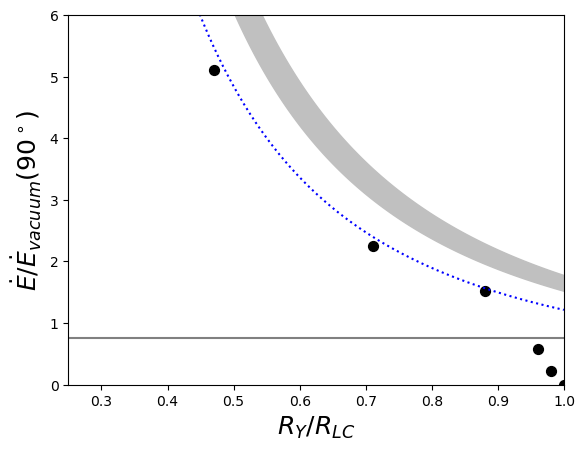}
\caption{Spindown rate $\dot{E}= \int_{\Psi=0}^{\Psi_{\rm open}}I(\Psi)\ {\rm d}\Psi\ \Omega/c$ in units of $\dot{E}_{\rm vacuum}(90^\circ)$ as function of the position of the Y-point $x_{\rm Y}$. Bullet points: solutions listed in Table~\ref{Table}. Dotted blue line: $\dot{E}=(1.10)^2\ \dot{E}_{\rm vacuum}(90^\circ)/x_{\rm Y}^2$. Grey band: canonical values $\dot{E}=(1.23-1.33)^2\ \dot{E}_{\rm vacuum}(90^\circ)/x_{\rm Y}^2\sim 1.5\ \dot{E}_{\rm vacuum}(90^\circ)/x_{\rm Y}^2$ according to~\cite{CKF99,Timokhin:2006,S06}, etc.
Grey line: the value $\dot{E}=0.75\ \dot{E}_{\rm vacuum}(90^\circ)$. As $x_{\rm Y}\rightarrow 1$, $\dot{E}\rightarrow 0$.}
\label{spindown}
\end{figure}

In Figure~\ref{Psiopen}, the variation in the open magnetic flux $\Psi_{\rm open}$ is plotted in units of $\Psi_{\rm dipole\ LC}$  with the position of the Y-point. Our present result that $\Psi_{\rm open}\rightarrow 0$ as $x_{\rm Y}\rightarrow 1$ is in tension with all previous solutions of the pulsar magnetosphere. The dotted blue line corresponds to the fit $1.1\  \Psi_{\rm open}/\Psi_{\rm dipole\ LC}/x_{\rm Y}$, while the grey band corresponds to the range of values obtained from the simulations of~\cite{Timokhin:2006}. For example, solution No~4 from Table~\ref{Table} contains the same amount of open magnetic flux as the one obtained by~\cite{hu2022axisymmetric}, yet their Y-point lies at $x_{\rm Y}=0.85$, while ours lies at $x_{\rm Y}=0.71$.
We believe that this pronounced difference is mainly due to our treatment of the separatrix current sheet as a contact discontinuity. In all previous solutions (e.g.,~\cite{CKF99,S06,Timokhin:2006,hu2022axisymmetric,hakobyan2023magnetic}), the separatrix contains a finite amount of poloidal magnetic flux and has a pronounced nonzero thickness. This affects the pressure balance across it which significantly modifies the position of the magnetospheric Y-point. The difference may also be partially due to our choice of a rather large stellar radius $r_*=R_{\rm LC}/4$ compared to previous solutions.
Finally, in Figure~\ref{spindown} the variation in the pulsar spindown energy loss $\dot{E}$ is plotted in units of $\dot{E}_{\rm vacuum}(90^\circ)$ with the position of the Y-point. Our present result that $\dot{E}\rightarrow 0$ as $x_{\rm Y}\rightarrow 1$ is {also} in tension with all previous solutions of the pulsar magnetosphere. The grey band corresponds to the spindown energy losses obtained from the simulations of~\cite{CKF99,contopoulos2005coughing,Timokhin:2006}. As is acknowledged above, the difference between the dotted line and the grey band may be partially due to our choice of a rather large stellar radius. The value shown with the horizontal line corresponds to the particular value $\dot{E}=0.75\ \dot{E}_{\rm vacuum}(90^\circ)$ mentioned in Equation~(\ref{intermittent}).

\unskip

\section{A Possible Connection with Fast Radio Bursts}\label{sec:4}

Time-dependent numerical simulations of the pulsar magnetosphere start with an initial configuration ${\bf B}(r,\theta,\phi ;t=0)={\bf B}_{\rm dipole}(r,\theta,\phi ;t=0)$ that corresponds to a magnetostatic dipole in vacuum, and ${\bf E}(r,\theta,\phi ;t=0)=0$ everywhere.
At time $t=0$, the programmer introduces electric fields ${\bf E}(r_*,\theta,\phi ;t)=-r\sin\theta\ \Omega/c\ \hat{\phi}\ \times {\bf B}_{\rm dipole}(r_*,\theta,\phi ;t)$ along the surface of the star, thus effectively setting the central star in rotation with angular velocity $\Omega$ along the $z$-axis. Here, $r_*$ is the radius of the star, and ${\bf B}_{\rm dipole}(r_*,\theta,\phi ;t)$ corresponds to a dipole field that rotates with angular velocity $\Omega$ along the $z$-axis. This is the procedure followed in the pioneering paper of Spitkovsky~\cite{S06} and in all time-dependent numerical simulations since then. The abrupt introduction of a non-zero electric field along the surface of the star generates a transient blast wave of poloidal electric and azimuthal magnetic fields that sweeps through the static magnetosphere. This is analogous to the transient blast wave of electric and magnetic fields generated when an electric charge is abruptly set in motion~\cite{Jackson:1975} 
or an electric current is introduced~\cite{Gourgouliatos:2008}. The aim of such simulations is to obtain the steady-state solution; thus, the programmer is not interested in the blast wave and waits for it to leave the inner magnetosphere. Indeed, after that wave sweeps through the computational grid that corresponds to the inner magnetosphere, a breathing steady-state solution is established with closed and open magnetic field lines. That initial artificial polarized blast wave has the characteristics of a fast radio burst (hereafter FRB).

FRBs represent one of the most intriguing enigmas in the field of astrophysics today. First discovered in 2007~\cite{Lorimer:2007}, FRBs are transient radio pulses that originate from distant galaxies, characterized by their extremely high energy and incredibly short duration, typically lasting just a few milliseconds. Despite their fleeting nature, these cosmic phenomena release more energy in a fraction of a second than the sun in an entire day.
The exact mechanisms that produce FRBs remain a subject of intense debate and speculation among scientists. Hypotheses range from highly magnetized neutron stars, known as magnetars, to more exotic theories involving cosmic strings or even extraterrestrial intelligence. However, despite over a decade of observation and research, no single theory has been able to fully explain all observed properties of FRBs.

We propose here that FRBs may be related to the electromagnetic burst of the toroidal magnetic field that follows an abrupt transition from the new solution that was obtained in this paper to one of the standard magnetospheric solutions of Figure~\ref{crazymag} and Table~\ref{Table}.
The duration of the burst will be on the order of the light crossing time of the light cylinder, namely $R_{\rm LC}/c=P/(2\pi)$, where $P$ is the spin period of the pulsar. 
The total energy of the burst is estimated to be equal to the difference between the electromagnetic energy of the new solution and the electromagnetic energy inside the light cylinder of a standard magnetospheric solution.  An estimate of the former is obtained as follows: 
according to Figures~\ref{crazymag} and \ref{Bz}, the dipolar magnetic flux $\Psi_{\rm dipole\ LC}$ that would have extended beyond the light cylinder, is now confined to close within an annular region of area $\sim R_{\rm LC}\delta$ where $\delta\sim 0.1 R_{\rm LC}$. That region has volume $\sim R_{\rm LC}^2 \delta\sim 0.1 R_{\rm LC}^3$, and in it, 
$B\sim \Psi_{\rm dipole\ LC}/(0.1 R_{\rm LC}^2) \sim 10 B_* r_*^3/R_{\rm LC}^3$. Therefore, 
\begin{equation}
E_{\rm new\ solution}\sim (10 B_* r_*^3/R_{\rm LC}^3)^2\ 0.1 R_{\rm LC}^3 \sim 10 B_*^2 r_*^6/R_{\rm LC}^3
\end{equation}

An estimate of the latter is 
\begin{equation}
E_{\rm standard\ solution}\sim (B_* r_*^3/R_{\rm LC}^3)^2 R_{\rm LC}^3 =B_*^2 r_*^6/
R_{\rm LC}^3
\end{equation}

Hence, the energy of the burst will be equal to the difference between the two, namely
\begin{eqnarray}
E_{\rm FRB} & \sim & E_{\rm new\ solution}-E_{\rm standard\ solution}\sim 
10 E_{\rm standard\ solution}\nonumber\\
& \sim &
10^{43}\left(\frac{B_*}{10^{13}\ {\rm G}}\right)^2 \left(\frac{P}{{\rm ms}}\right)^{-1}\ {\rm erg}
\end{eqnarray}

Notice that in the energy estimates of this section, we use a physical neutron star radius of $r_*\approx 10$~km.

As we will now see, this is more than enough to power an FRB. In the case of an accreting recycled millisecond pulsar with a surface magnetic field on the order of $B_*\sim 10^{8-9}$~G, 
\begin{equation}
E_{\rm FRB\ ms\ pulsar} \sim
10^{35}\left(\frac{B_*}{10^{9}\ {\rm G}}\right)^2 \ {\rm erg}\sim L_{\odot}\times 100\ {\rm s}\ \left(\frac{B_*}{10^{9}\ {\rm G}}\right)^2\ ,
\end{equation}
where $L_\odot$ is the solar luminosity. This value is below the lowest limit of the low-energy FRBs presented by~\cite{Gourdji:2019}; hence, it does not seem consistent with actual FRB observations.  {For a newborn neutron star with a millisecond period formed via a core collapse supernova,}  
\begin{equation}
E_{\rm FRB\ newly\ formed\ ns} \sim
L_{\odot}\times 100\ {\rm yr}\ \left(\frac{B_*}{10^{13}\ {\rm G}}\right)^2\ .
\end{equation}

Actual FRB energies lie between the above extreme limits, namely between $10^{37}$ and $10^{41}$~erg (e.g.,~\cite{FRB2,FRBCatalogue}). Therefore, we propose that FRBs correspond to millisecond pulsars with surface magnetic fields between $10^{11}$ and $10^{12}$~G.

\section{Discussion and Conclusions}\label{sec:5}

We have improved the solutions of the pulsar equation when the tip of the closed-line region, the so-called Y-point, approaches the light cylinder. The new solutions were obtained under the assumption that the separatrix surface between open and closed field lines is a zero-thickness contact discontinuity. The new solutions are similar to the old ones of~\cite{CKF99,contopoulos2005coughing,Timokhin:2006} when the Y-point lies inside about $80\%$ of the light cylinder radius. As the Y-point approaches closer to the light cylinder, however, the solutions diverge from the standard ones found in the literature, and in the extreme limit that the closed-line region touches the light cylinder, the open-line region disappears completely. As is seen in Equation~(\ref{pressurebalance}) in Appendix~\ref{app:level2} and in Figure~\ref{Bz}, the closer the Y-point to the light cylinder, the stronger the divergence of ${\rm B}_p$ right inside the Y-point at the tip of the closed-line region needs to be. We see no other way to implement this divergence of ${\rm B}_p(x_{\rm Y}\rightarrow 1^-)$ except with more and more field lines to enter the closed-line region and compress the magnetic field at its tip. This effect is clearly seen in the sequence of solutions shown in Figure~\ref{solutions}. In the limit where ${\rm B}_p\rightarrow \infty$ as $x_{\rm Y}=1^-$, all field lines enter the closed-line region confined inside the light cylinder, and the open line region disappears completely (Figure~\ref{crazymag}).

One may legitimately ask why these solutions were never before obtained in the literature. There are several reasons. Firstly, in all previous solutions, the separatrix surface along which the magnetospheric electric current circuit closes on the star had non-zero thickness due to the finite resolution of the code. In particular, at its tip, the outer part of the separatrix forms a Y-point that touches the light cylinder, while the inner part forms a clear T-point some distance inside it. This is clearly explained in Figure~4 of~\cite{Contopoulos:2024}. Unfortunately, the ideal force-free conditions of Equation~(\ref{pulsareq1}) are not valid inside current sheets (as, e.g., in Figure~1b of~\cite{S06}). Nevertheless, the fact that PIC simulations also form a separatrix that contains a nonzero amount of poloidal magnetic flux (see high-resolution detail in Figure~9 of~\cite{hu2022axisymmetric}) led us to consider the possibility that this effect is indeed physical. It is interesting that the separatrix current sheet contains a nonzero guide field and is thus different from the equatorial current sheet which is an ordinary Harris-type current sheet with $180^\circ$ field inversion across it. It also seems that the width of the former type of current sheet grows with time as is seen in Figure~1c of~\cite{S06}. This effect merits further investigation with hybrid PIC-FFE simulations  {(Soudais, Cerutti \& Contopoulos~2024, submitted)}
. 
Now that the separatrix has zero thickness, it forms a clear T-point at its tip, which indeed lies some distance inside the light cylinder. Secondly, time-dependent simulations always start from a vacuum dipole configuration that extends to all space and therefore, some fraction of it already crosses the light cylinder. After the central star is set into rotation, the magnetosphere is divided into open and closed field lines, i.e., there is no chance to obtain a configuration totally enclosed inside the light cylinder as the solution shown in Figure~\ref{crazymag}. Thirdly, after the time-dependent evolution relaxes to a steady-state solution, it is reasonable to expect that it will further relax to the lowest energy solution. Therefore, time-dependent simulations cannot yield the full sequence of solutions up to the latter singular solution as our Machine Learning methodology can. It is interesting that, now that the full sequence of solutions is obtained,  it is shown that the solution where $x_{\rm Y}\sim 80\%\ R_{\rm LC}$ is a minimum energy solution, and this is why previous time-dependent high-resolution PIC simulations relax to that solution. We have thus improved upon the result of~\cite{Contopoulos:2024} which placed the minimum energy solution around  $x_{\rm Y}=92\%\ R_{\rm LC}$.

We argue that the confined singular solution of Figure~\ref{crazymag} and all solutions between that one and the minimum energy solution are unstable; thus, if the magnetospheric configuration of Figure~\ref{crazymag} forms during a supernova explosion 
{
or during a high accretion event from a surrounding disk or wind from a donor star,
}
it will transition to the minimum energy solution by emitting a burst of energy equal to the energy difference between these two solutions. We propose that this burst may be related to an FRB. 
 {We would like to notice here that, as long as the surrounding material is there, the confined solution cannot transition to a standard solution in which part of the magnetosphere extends to infinity, and thus the accumulated electromagnetic energy of the confined solution cannot be released. What happens when the surrounding material dissolves and how fast electromagnetic energy is released remain to be seen in a time-dependent numerical simulation of this transition.} 
In case the pulsar magnetosphere undergoes a later accretion event from an external disk or wind from a donor star that will re-confine it to the solution of Figure~\ref{crazymag}, the FRB will repeat. 
{
We must acknowledge of course that a few FRBs have already been observed from a magnetar, and that the cosmological sky rate of FRBs exceeds that of core-collapse supernovae; therefore, our model may just be a rare kind of FRB, if feasible at all. Moreover, we have not answered why would the emission of this state transition be in the MHz to GHz radio regime, and how it would escape the very dense accretion environment that formed the confined solution of Figure~\ref{crazymag}.
Our model certainly merits further investigation with detailed MHD numerical simulations.
}

{
What is also interesting to investigate is whether a confined solution exists for non-aligned (oblique) pulsars in which $\partial{\bf B}/\partial t\neq {\bf 0}$. We expect that oblique rotators would radiate electromagnetic radiation to infinity; thus, magnetic field lines would extend to infinity. Therefore, it seems impossible for an oblique magnetosphere to be totally confined inside the light cylinder. It may thus be that the confined solution of the type shown in  {Figure}~\ref{crazymag} is only possible in aligned pulsars. 
}  

\authorcontributions{Conceptualization and supervision, I.C.; methodology, I.C. and K.N.G.; software and visualization, D.N. and I.D.;  
writing—original draft preparation, I.C.; writing—review and editing, all authors.
All authors have read and agreed to the published version of the manuscript.}

%
\funding{This research work was partially funded by the Hellenic Foundation for Research and Innovation (HFRI).}

%
\dataavailability{The datasets generated during the current study are available from the corresponding author upon reasonable request.}

%
\acknowledgments{ {We would like to thank G. Contopoulos and A. Nathanail for interesting discussions on the origin of FRBs. We would also like to thank the International Space Science Institute (ISSI) for providing financial support for the organization of the meeting of ISSI Team No 459 led by I. Contopoulos and D. Kazanas where the ideas presented in the paper originated.
This research work was supported by the Hellenic Foundation for Research and Innovation (HFRI) under the 4th Call for HFRI PhD Fellowships (Fellowship Number: 9239). 
This work was supported by computational time granted from the National Infrastructures for Research and Technology S.A. (GRNET S.A.) in the National HPC facility - ARIS - under project ID pr015026/simnstar.}} 

\conflictsofinterest{The authors declare no conflicts of interest.}

\appendixtitles{no} 
\appendixstart
\appendix
\section[\appendixname~\thesection]{} \label{app:level1}

In the first paper of this series (\cite{Dimitropoulos:2024}, Paper~I), a new methodology was presented for solving the pulsar equation which essentially consists of the following steps:
\begin{enumerate}
\item The angular size $\theta_{\rm pc}$ of the polar cap that contains the so-called open magnetic field line is chosen a priori. The rest of the stellar surface contains closed field lines.
\item The two regions of open and closed magnetic field lines are considered independently. An `infinitely thin' physical discontinuity exists between the two regions in the form of  a separatrix current sheet. In reality, the separatrix current sheet has some finite thickness which is orders of magnitude smaller than the characteristic dimensions of the system such as the radius of the star and the radius of the light cylinder. 
\item  It is very helpful to deal with the equatorial current sheet in the open line region by mathematically reversing the polarity of the magnetic field that emanates from the south pole of the star. By carrying this out, the mathematical discontinuity in the equatorial region disappears; thus, one does not need to worry about it computationally. At the end of the calculation, we will return to the original polarity of the magnetic field, and the equatorial current sheet will reappear.
\item The force-free problem is solved in the open and closed line regions independently. Our method of choice is with Neural Networks so-called Physics Inspired (PINNs). One may also use classical grid methods in the two regions.
\item The shape $r_{\rm S}(\theta)$ of the separatrix is readjusted so that continuity of ${\rm B}^2-{\rm E}^2$ is achieved at all points across it. One way to carry that out is to move the separatrix radially at each point with an amount that is proportional to the difference $({\rm B}^2-{\rm E}^2)_{\rm IN}-({\rm B}^2-{\rm E}^2)_{\rm OUT}$.
\item Repeat the procedure until pressure balance is achieved over the whole separatrix and its shape stabilizes.
\end{enumerate}

{
In~\cite{Dimitropoulos:2024}, Machine Learning was implemented in the PyTorch Deep Learning library. Code development was implemented in Python on Anaconda Jupyter Notebooks running on local GPUs, while production runs were performed in the Cloud in Google’s Colab. Furthermore, Adam optimizers, SiLU activation functions, and ReduceLROnPlateau schedulers from the PyTorch library were used. Three independent NNs were trained:
\begin{enumerate}
\item One PINN with two entries (the $r,\theta$ spherical coordinates), three hidden layers with 64 nodes each, and two exits $\Psi,I$. This PINN solves the pulsar equation in the open line region outside the separatrix.
\item One PINN with two entries $r,\theta$, three hidden layers with 64 nodes each, and one exit $\Psi$. This PINN solves the pulsar equation in the closed line region inside the separatrix where I = 0.
\item A third NN with one entry $\theta$, two hidden layers with 128 nodes each, and one exit $r_{\rm S}$ is trained to yield the shape $r_{\rm S}(\theta)$ of the separatrix at all angles $\theta$.
\end{enumerate}
The number of internal nodes and layers in the two PINNs that solve the pulsar equation was chosen by trial and error to be able to reproduce known features of the solution. In practice, the two main PINNs are initially trained for 50,000 steps after which, satisfactory convergence is achieved in the two regions. After that initial training stage, the separatrix is displaced based on the resulting pressure differences between the closed and open line regions, and then the third NN is trained. The two main PINNs are re-trained for another 50,000 steps, and the process is repeated 10 times. In total, we run 500,000 training steps for the two main PINNs. In the resulting configuration, pressure balance is achieved across the separatrix to less than $1\%$. 
}

As it is acknowledged in~\cite{Dimitropoulos:2024}, the numerical accuracy of our method is not on par yet with that of standard methods such as finite difference, finite volume or spectral methods. This is also the case in the solution of~\cite{stefanou:2023}. On the other hand, the numerical treatment of current sheets by standard methods it too is problematic (numerical current sheets have unphysical thicknesses and the numerical dissipation there is questionable). In order to improve our results in future iterations of this work, we will try to combine standard methods outside current sheets with our methodology of moving the separatrix. This is still a work in progress. Modulo the above disclaimers, we applied this method to obtain the sequence of solutions shown in Figure~\ref{solutions}.

\section[\appendixname~\thesection]{} \label{app:level2}

We discuss here the shape of the tip of the closed-line region, the so-called Y-point. In Paper~I, it was found that
\begin{enumerate}
\item The Y-point is a clear T-point as predicted by~\cite{uzdensky2003axisymmetric}, not a Y-point as is seen in all numerical solutions of the axisymmetric pulsar magnetosphere to date, except for~\cite{Contopoulos:2024}. This is probably due to the fact that in all previous numerical simulations, the separatrix between closed and open field lines had an unphysically large thickness (sometimes on the order of the light cylinder!); thus, it was not a true contact discontinuity as in Paper~I.
\item The Y-point of our first solution obtained in Paper~I was found to lie closer to the star than the corresponding Y-point of the standard FFE solution obtained previously. 
\end{enumerate}

Both of the above results modify our understanding of how fast pulsars spin down. 
In the standard solution, the two regions are separated by a separatrix surface that contains a charged current sheet. One very interesting point in understanding the operation of the pulsar magnetosphere is how close to the light cylinder can the Y-point lie. According to Equation~(\ref{pressurebalance1}), the pressure balance condition across the separatrix current sheet may be written as
\begin{equation}
\left. {\rm B}_p^2(1-x^2)\right|_{\rm IN}=\left. {\rm B}_p^2(1-x^2)\right|_{\rm OUT}+\frac{I(\Psi_{\rm S})^2}{x^2}
\label{pressurebalance}
\end{equation}

Here, $I(\Psi_{\rm S})$ is the electric return current flowing along the separatrix, and $p$-indices denote poloidal ($r,\theta$) components of the fields. Let us first repeat the main argument why the tip of the closed line region on the equator is a T-point. The poloidal magnetic field $\left. {\rm B}_p\right|_{\rm OUT}$ in the right hand side of Equation~(\ref{pressurebalance}) is obviously equal to zero right above and below the equatorial current sheet. The third term, however, is finite because of the presence of the nonzero equatorial current $2I(\Psi_{\rm S})$ (notice that the equatorial current sheet contains double the current of each separatrix current sheet). Therefore, the only way to satisfy Equation~(\ref{pressurebalance}) is to have $\left. {\rm B}_p\right|_{\rm IN}\neq 0$. This can only be satisfied inside a T-point, not inside a Y-point. Nevertheless, we will continue calling the tip of the closed-line region the Y-point because of its overall bifurcating shape, and we will denote the rescaled radial distance of the tip of the closed-line region with $x_{\rm Y}$.

Furthermore, it is obvious that Equation~(\ref{pressurebalance}) becomes singular when $x_{\rm Y}$ approaches very close to unity. In that limit, Equation~(\ref{pressurebalance}) yields
\begin{equation}
\left. {\rm B}_p(x_{\rm Y})\right|_{\rm IN}=
\frac{I(\Psi_{\rm S})}{x_{\rm Y}\sqrt{1-x_{\rm Y}^2}}=
\frac{I(\Psi_{\rm S})}{\sqrt{2}\sqrt{1-x_{\rm Y}}}\rightarrow\infty\ \ \ \mbox{when}\ \ x_{\rm Y}\rightarrow 1\ .
\label{div}
\end{equation}

As is shown in this paper, contrary to what was obtained in previous works, it is not possible to reach this limit and have a significant open-line region outside. The closer the Y-point approaches the light cylinder, the more field lines must be squeezed inside the closed-line region to support the growth of ${\rm B}_p$ at its tip required by Equation~(\ref{div}), the fewer field lines are left in the open-line region. As is shown with the new confined magnetospheric solution of Figure~\ref{crazymag}, in the limit $x_{\rm Y}\rightarrow 1^-$ the open-line region disappears completely.

\begin{adjustwidth}{-\extralength}{0cm}
\reftitle{References}

\externalbibliography{yes}

\PublishersNote{}
\end{adjustwidth}

\end{document}